\documentclass[4apaper,final,mathptmx,cmfonts,numberedheadings]{aipproc}
\layoutstyle{6x9}

\usepackage{amssymb,amsmath}

\def\noi{\noindent}
\def\bea{\begin{eqnarray}}
\def\eea{\end{eqnarray}}
\def\nn{\nonumber}
\def\vsj{\vspace{1.5mm}}
\def\vsp{\vspace{.5mm}}

\def\vsi{\vspace{.5em}}
\def\vss{\vspace{1.5em}}
\def\vsa{\vspace{2em}}
\def\vsr{\vspace{2.5em}}

\begin{document}

\title{$p$-Adic and Adelic Cosmology: \\  $p$-Adic Origin of Dark Energy and Dark Matter}


\author{Branko  Dragovich}{
  address={Institute of Physics, P.O.Box 57, 11001 Belgrade, SERBIA AND MONTENEGRO\\
  email: \emph{dragovich@phy.bg.ac.yu}}
}

\begin{abstract}
 A brief review of $p$-adic and adelic cosmology is presented.
 In particular, $p$-adic and adelic aspects of gravity, classical
 cosmology, quantum mechanics, quantum cosmology and the
 wave function of the universe are considered.\vsp

\noi  $p$-Adic worlds made of $p$-adic matters, which are different
from
 real world of ordinary matter, are introduced. Real world and
 $p$-adic worlds make the universe as a whole. $p$-Adic origin of
 the dark energy and dark matter are proposed and discussed.
\end{abstract}

\maketitle

%%%%%%%%%%%%%%%%%%%%%%%%%%%%%%%%%%%%%%%%%%%%
%% MAINMATTER
%%%%%%%%%%%%%%%%%%%%%%%%%%%%%%%%%%%%%%%%%%%%

\vsi

\section{Introduction}

Cosmology in its own right is a science devoted to the universe as a
whole. It is based on the cosmological observational data and
fundamental physical theories (especially: general theory of
relativity, quantum theory and theory of elementary particles). So
far many significant results have been obtained and the Standard
Cosmological Model is established (see, e.g. \cite{scott}).
According to this model at the very beginning the universe was very
small, dense, hot and started to expand. This initial period of
evolution should be described by quantum cosmology. According to
general theory of relativity the universe is a (pseudo)Riemannian
space, which is presently flat. From observational data follows that
the universe has permanently expanded during all its history which
is about 14 billion years. Since 1998, there are a lot of evidence
that the universe is now in the stage of an accelerated expansion
which began a few  billion years ago (for a review, see
\cite{sahni1,padmanabhan1}). To explain this acceleration many
models have been proposed but no one of them was generally accepted.
A very natural and the most attractive is the approach with dark
energy, which is a matter with a negative pressure and uniformly
distributed in the space. About $70 \%$ of all energy content of the
universe is related to the dark energy, while about $26 \%$ belongs
to the dark matter and only $4 \%$ is made of baryons (protons and
neutrons). However the nature of the dark energy is a big mystery,
which presents one of the greatest problems and challenges of
contemporary cosmology and theory of elementary particles.\vsj

\noi It is well known that results of  observational measurements,
as well as of the experimental ones, are elements of the field
${\mathbb{Q}}$ of rational numbers. However theoretical models are
constructed not over  $\mathbb{Q}$ but traditionally over
$\mathbb{R}$ (field of real numbers) or $\mathbb{C}$ (field of
complex numbers), where $\mathbb{R}$ is completion of $\mathbb{Q}$
with respect to distance induced by the ordinary absolute value
$|\cdot|_\infty$ and $\mathbb{C}$ is algebraic extension of
$\mathbb{R}$. Thus it is worth noting that mathematical modeling of
physical systems requires not only algebraic structure of
${\mathbb{Q}}$ but also its geometric properties, which are related
to  the possible norms on ${\mathbb{Q}}$. Besides $|\cdot|_\infty$
there are also infinitely many $p$-adic norms $|\cdot|_p$ , i.e.
there is one nontrivial and inequivalent norm for each prime number
$p$ \cite{schikhof,gelfand}.\vsj

\noi Measurements represent comparison of a given quantity with
respect to a fixed quantity of the same nature taken to be the unit
one. We live in the world where our measurements are inherently
connected with the (decimal) expansions of the real numbers. Namely,
in the process of measurement one determines a finite number of
digits in the decimal expansion. This number of digits can be
enlarged using more precise tools, while all other digits remain
hidden within measuring errors. According to the classical mechanics
there are no in principle physical restrictions to measure all
quantities of a system with arbitrary accuracy. However in quantum
mechanics there is restriction as a consequence of the well known
uncertainty relation $\Delta x\, \Delta k \geq \hbar/2$. When
quantum mechanics is combined with gravity then there obtains strong
restriction on measurement of very small distances in the form
\begin{equation}
 \Delta x \, \geq \,  \ell_0 \, = \, \sqrt{\frac{G\, \hbar}{c^3}} \, \sim \,
 10^{-33}\,   \mbox{cm}  \, ,  \label{1.1}
\end{equation}
where $\ell_0$ is the Planck length. It follows that there is quite
firm  restriction to measure spatial distances with arbitrary
accuracy. In other words, it is not possible to determine all digits
related to positions in the space. Thus measurements of physical
quantities  up to the Planck scale give rational numbers with
geometrical properties which can be described  by the usual absolute
value. Hence it has been natural to use real numbers and complex
numbers to describe physical phenomena  in the explored domain of
the universe. It is not strange that just real analysis is used to
describe these real data. However for a more profound physical
theory  there is a sense to employ also $p$-adic numbers
\cite{schikhof,vladimirov1}, which are
 completions of ${\mathbb{Q}}$  with respect to the $p$-adic
norms.\vsj

\noi If we cannot get as a result of direct measurement a rational
number with $p$-adic norm properties, it does not mean that there is
not any content of the universe which natural description is just by
$p$-adic numbers. Suppose that such  $p$-adic systems exist. Then
real number, as a result of measurement, is a real measure  of
interaction between $p$-adic  and real system, to which belong
measuring instruments. In such case we have to use analysis  with
$p$-adic  valued functions of $p$-adic argument to describe $p$-adic
system itself and also to use analysis with real (complex) valued
functions of $p$-adic argument to describe  $p$-adic  system from
the real point of view. This is slightly similar to the employment
of complex functions in quantum mechanics, where wave
function\linebreak $\psi (x) \in \mathbb{C}$ contains complete
information on quantum system, but is not measurable quantity, while
$|\psi (x)|^2 \in \mathbb{R}$ is related to the probability
distribution which can be measured. Let us use terms real and
$p$-adic to denote those aspects of the universe which can be
naturally described by real and $p$-adic numbers, respectively. We
conjecture here that the visible and dark sides of the universe are
real and $p$-adic ones, respectively.\vsj

\noi $p$-Adic strings were introduced in 1987 \cite{volovich1} and a
nice adelic formula was obtained \cite{freund1}. An effective theory
with real numbers of $p$-adic strings  was constructed
\cite{brekke1,frampton1} and shown  its importance in the context of
the tachyon condensation \cite{sen1}. Application of $p$-adic
numbers in construction of various models for the first five years
was presented in \cite{brekke2} and \cite{vladimirov1}. Current
activity is reflected in the proceedings of the international
conferences on $p$-adic mathematical physics \cite{conf1,conf2}.\vsj

\noi In Section 2 we give an introductory review of adeles as well
as of $p$-adic and adelic analogs of classical cosmology. Sec. 3 is
devoted to $p$-adic and adelic quantum mechanics and their
employment in quantum cosmology. $p$-Adic matter and its
cosmological aspects are considered in Sec. 4. \vss

\section{{\it\lowercase{$\textbf{p}$}}-Adic and Adelic Classical Cosmology}

Adelic classical cosmology is a generalization of the ordinary one
(over real numbers) in  such way that it employs all (real and
$p$-adic) completions of the set of rational numbers $\mathbb{Q}$.
It uses analysis based on adelic valued functions of adelic valued
arguments.\vsj

\noi Let us remind some basic properties of adeles and adelic valued
functions. To consider real and $p$-adic numbers simultaneously and
on equal footing one uses concept of adeles. An adele $x$ (see, e.g.
\cite{gelfand}) is an infinite sequence
\begin{equation}
  x= (x_\infty, x_2, \cdots, x_p, \cdots)\,,  \quad
x_\infty \in \mathbb{R} \,, \,\,\, x_p \in \mathbb{Q}_p \label{2.1}
 \end{equation}
 with the restriction that for all but a finite set $\mathcal P$ of
primes $p$ must be $x_p \in {\mathbb Z}_p $, where ${\mathbb Z}_p
= \{ y \in \mathbb{Q}_p \, | \, |y|_p \leq 1 \}$ is the ring of
$p$-adic integers. Componentwise addition and multiplication endow
the ring structure to the set of all adeles ${\mathbb A}$ , which
is the union of  direct products in the following form:
\begin{equation}
 {\mathbb A} = \bigcup_{{\mathcal P}} {\mathbb A} ({\mathcal P}),
 \ \ \ \  {\mathbb A} ({\mathcal P}) = {\mathbb R}\times \prod_{p\in
 {\mathcal P}} {\mathbb Q}_p
 \times \prod_{p\not\in {\mathcal P}} {\mathbb Z}_p \, .         \label{2.2}
\end{equation}

\noi A multiplicative group of ideles ${\mathbb I}$ is a subset of
${\mathbb A}$ with elements $x= (x_\infty, x_2, \cdots, x_p,
\cdots)$,  where $x_\infty \in {\mathbb R}^\ast = {\mathbb R}
\setminus \{ 0\}$ and $x_p \in {\mathbb Q}^\ast_p = {\mathbb Q}_p
\setminus \{0 \}$ with the restriction that for all but a finite set
$\mathcal P$  one has   $x_p \in {\mathbb U}_p$, where ${\mathbb
U}_p = \{ y \in \mathbb{Q}_p \, | \, |y|_p = 1 \}$ is the
multiplicative group of $p$-adic units . Thus the whole set of
ideles is
\begin{equation}
 {\mathbb I} = \bigcup_{{\mathcal P}} {\mathbb I} ({\mathcal P}),
 \ \ \ \ {\mathbb I} ({\mathcal P}) = {\mathbb R}^{\ast}\times \prod_{p\in {\mathcal P}}
 {\mathbb Q}^\ast_p
 \times \prod_{p\not\in {\mathcal P}} {\mathbb U}_p \, .     \label{2.3}
\end{equation}

\noi A principal adele (idele) is a sequence $ (x, x, \cdots, x,
\cdots) \in {\mathbb A}$ , where $x \in  {\mathbb Q}\quad (x \in
{\mathbb Q}^\ast = {\mathbb Q}\setminus \{ 0\})$. ${\mathbb Q}$ and
${\mathbb Q}^\ast$ are naturally embedded in  ${\mathbb A}$ and
${\mathbb I}$ , respectively.\vsj

\noi Let us define an ordering on the set ${\mathbb P}$, which
consists of all finite sets ${{\mathcal P}_i}$ of primes $p$, by
${{\mathcal P}_1} \prec {\mathcal P}_2$ if ${{\mathcal P}_1} \subset
{\mathcal P}_2$. It is evident that ${\mathbb A}({\cal P}_1)\subset
{\mathbb A}({\cal P}_2)$ when ${\mathcal P}_1 \prec {\mathcal P}_2$.
Spaces ${\mathbb A}({\cal P})$ have natural Tikhonov topology and
adelic topology in ${\mathbb A}$ is introduced by inductive limit: $
{\mathbb A} = \lim \mbox{ind}_{{\mathcal P} \in {\mathbb P}}
{\mathbb A}({\cal P})$. A basis of adelic topology is a collection
of open sets of the form $ W ({\mathcal P}) = {\mathbb V}_\infty
\times \prod_{p \in {\mathcal P}} {\mathbb V}_p\, \times \prod_{p
\not \in {\mathcal P}} {\mathbb Z}_p \, $, where ${\mathbb
V}_\infty$ and ${\mathbb V}_p$ are open sets in ${\mathbb R}$ and
${\mathbb Q}_p$ , respectively. Note that adelic topology is finer
than the corresponding Tikhonov topology. A sequence of adeles
$a^{(n)}\in {\mathbb A}$ converges to an adele $a \in {\mathbb A}$
if ${(\it i})$ it converges to $a$ componentwise and $({\it ii})\, $
if there exist a positive integer $N$ and a set ${\mathcal P}$ such
that $\, a^{(n)}, \, a \in {\mathbb A} ({\mathcal P})$ when $n\geq
N$. In the analogous way, these assertions hold also for idelic
spaces ${\mathbb I}({\cal P})$ and ${\mathbb I}$.\vsj

\noi Adelic valued functions of adelic arguments are maps
$F_{\mathbb{A}} : U_{\mathbb{A}} \to V_{\mathbb{A}}$, where
$U_{\mathbb{A}} \subset \mathbb{A}^n \,,$ $ \,\,\, V_{\mathbb{A}}
\subset \mathbb{A}^m $ and have the form
\begin{equation}
F_{\mathbb{A}} (x) = (f_\infty (x_\infty)\,, \, f_2 (x_2)\,, \cdots
,\, f_p (x_p)\,, \cdots ) \,,   \quad f_\infty \in \mathbb{R}^m \,,
\, \,\,\,  f_p  \in  \mathbb{Q}^m_p \,, \label{2.4}
\end{equation}
where for all but $p \in \mathcal{P}$ one has to satisfy $| f_p
(x_p)|_p \leq 1$.\vsj

\noi When $F_{\mathbb{A}} (x) $ is related to the same physical
system it is natural to expect that $v$-adic $(v = \infty, 2,
\cdots, p, \cdots)$ functions $f_v (x_v)$ have the same form of
dependence on $x_v$. As an illustrative example one can take adelic
valued exponential function
\begin{equation}
\exp_{\mathbb{A}} x = (\exp_\infty\, x_\infty \,, \exp_2\, x_2 \,,
\cdots , \exp_p\, x_p \,, \cdots ) \,,  \label{2.5}
\end{equation}
 where $v$-adic exponential functions are defined by the
usual power series expansion
\begin{equation}
\exp_v \, x_v =  \sum_{n=0}^\infty  \frac{x_v^n}{n!} \,, \quad
|x_p|_p \leq |2 p|_p \,, \quad  |\exp_p\, x_p|_p = 1 .  \label{2.6}
\end{equation}
Similar situation is for functions $ \sin_{\mathbb{A}} x \,,
\cos_{\mathbb{A}} x \,, \sinh_{\mathbb{A}} x \,, \cosh_{\mathbb{A}}
x$ and many other functions given by power series expansions with
rational coefficients.\vsj

\noi The Einstein gravitational field equations
\begin{equation}
R_{\mu \nu} - \frac{1}{2}\, R \, g_{\mu \nu}  = \kappa \, T_{\mu
\nu}  - \Lambda \, g_{\mu \nu}    \label{2.7}
\end{equation}
can be also considered as $p$-adic ones  if we take constants
$\kappa = \frac{8 \pi G}{c^4}$  and $\Lambda$ to be rational
numbers. By this way the Einstein equations \eqref{2.7} become
number field invariant and therefore  more fundamental. A successful
systematic study of $p$-adic gravity, especially $p$-adic
differential geometry and gravitational field equations, started in
1991 \cite{arefeva1}.\vsj

\noi For the sequel it is useful to introduce $ \pi \, G = \bar{G}$
and the new Planck quantities
\begin{equation}
L_0 = \sqrt{\frac{h \bar{G}}{c^3}} = \sqrt{2}\, \pi \,\ell_0 \,,
\quad T_0 = \frac{L_0}{c}=  \sqrt{2}\, \pi \, t_0 \,, \quad  M_0 =
\sqrt{\frac{h c}{\bar{G}}} = \sqrt{2}\,  m_0 \,, \label{2.8}
\end{equation}
where $\ell_0 \,, t_0 $  and $m_0$ are usual Planck length, time and
mass, respectively. Let us take for the natural system of units
these $L_0 \,, T_0$ and $M_0$ instead of the standard $\ell_0 \,,
t_0 $ and $m_0$. Then $\bar{G} \,, c$ and $h$ are rational numbers
with unit values.\vsj

\noi Constructing cosmological models it is useful to maximally
exploit symmetrical properties of the universe. As a consequence one
obtains dynamical system called minisuperspace cosmological model,
which contains finite number degrees of freedom. Minisuperspace
model may be regarded as a classical system given by Hamiltonian $H
(q^i \,, k^i \,, t)$  or Lagrangian $L (q^i \,, \dot{q}^i \,, t)$ ,
where $i = 1, 2, \cdots, n$. The corresponding adelic Hamiltonian is
\begin{equation}
H_{\mathbb{A}} (q^i \,, k^i \,, t) = (H_\infty (q^i_\infty \,,
k^i_\infty \,, t_\infty)\,,  H_2 (q^i_2 \,, k^i_2 \,, t_2) \,,
\cdots , H_p (q^i_p \,, k^i_p \,, t_p)\,, \cdots ) \,, \label{2.9}
\end{equation}
where $H_\infty \in \mathbb{R}$ and $H_p \in \mathbb{Q}_p$ with
restriction that $H_p \in \mathbb{Z}_p$ for all but $p \in
\mathcal{P}$ . Analogously one defines an adelic Lagrangian.
$v$-Adic action of the minisuperspace model is
\begin{equation}
S_v [q]  = \int_{t'}^{t''} dt\, N \Big[ \frac{1}{2 N^2} \, f_{\alpha
\beta} (q) \dot{q}^\alpha \, \dot{q}^\beta - U (q)\Big]\,,
\label{2.10}
\end{equation}
where $f_{\alpha \beta}$ is a metric on minisuperspace of some
gravitational and matter field variables.\vsj

\noi The de Sitter minisuperspace cosmological model is a simple
nontrivial, exactly solvable and instructive cosmological model. It
is given by the Einstein-Hilbert action with cosmological constant
$\Lambda$  $\, (c=1)$,
\begin{equation}
S [g_{\mu \nu}] = \frac{1}{16 \pi G} \Big[ \int_M d^4x \sqrt{- g} (R
-  2 \Lambda)  + 2 \int_{\partial M} d^3x \sqrt{h} K \Big]
\label{2.11}
\end{equation}
and the Friedmann-Robertson-Walker (FRW) metric
\begin{equation}
ds^2 =  -N^2 dt^2  + a^2(t) d\Omega^2_3\,,   \label{2.12}
\end{equation}
where $N(t)$ is the lapse function and $d\Omega^2_3$ is the metric
on the unit three-sphere. More complex models contain also
additional action with matter fields. To simplify formalism and get
quadratic Lagrangian one can take the Robertson-Walker metric in the
form \cite{halliwell}
\begin{equation}
ds^2 = -\frac{N^2}{q(t)} dt^2  + q(t) d\Omega^2_3 \,. \label{2.13}
\end{equation}
Using metric (\ref{2.13})  in (\ref{2.11}), one obtains
\begin{equation}
S [q] = \frac{1}{2}  \int_{t'}^{t''} dt \, N \Big( -
\frac{\dot{q}^2}{ 4 N^2} - \lambda q + 1 \Big) \,, \label{2.14}
\end{equation}
where $\lambda = \Lambda  /3$. Choosing $N = 1$, the equation of
motion is
\begin{equation}
\ddot{q} = 2 \lambda  .     \label{2.15}
\end{equation}
Solution of the equation (\ref{2.15}) which satisfies conditions
$q'' = q (T)$ and $q' = q (0)$  is
\begin{equation}
q(t) =  \lambda t^2  + \Big( \frac{q'' - q'}{T} - \lambda T  \Big) t
+ q' \, .   \label{2.16}
\end{equation}
Note that equations (\ref{2.15}) and (\ref{2.16}) are the same as
for a particle with constant acceleration $a = 2 \lambda$.  The
corresponding classical action (see, e.g. \cite{dragovich1} and
references therein) is
\begin{equation}
 \bar{S} [q]  = \frac{\lambda^2 T^3}{24} - [\lambda (q'' + q') - 2]
 \frac{T}{4} - \frac{(q'' - q')^2}{8 T} . \label{2.17}
\end{equation}

\noi The above consideration of the de Sitter model is performed for
the real case but the expressions from (\ref{2.14}) to (\ref{2.17})
can be also regarded as $p$-adic valued.\vsj

\noi Since parameter $\lambda = \frac{\Lambda}{3}$ has rational
values we can write adelic Lagrangian in the form
\begin{equation}
L_{\mathbb{A}} (q, \dot{q}) = (L_\infty (q_\infty ,
\dot{q}_\infty)\,, L_2 (q_2 , \dot{q}_2)\,, \cdots , L_p (q_p ,
\dot{q}_p) \,, \cdots) \,,  \label{2.18}
\end{equation}
where ($N=1$)
\begin{equation}
L_v (q_v , \dot{q}_v) = \frac{1}{2}  \,  \Big( - \frac{\dot{q}^2_v}{
4 } - \lambda q_v + 1 \Big) \,. \label{2.19}
\end{equation}
It is evident that $ |L_p (q_p , \dot{q}_p)|_p \leq 1$ when $|q_p|_p
\leq 1 \,, \,\, |\dot{q}_p|_p \leq 1 \,, \,\, |\lambda|_p \leq 1$
and $p \neq 2$. Infinite sequence of actions
\begin{equation}
\bar{S}_{\mathbb{A}} [q] = (\bar{S}_\infty [q_\infty] \,, \bar{S}_2
[q_2] \,, \cdots , \bar{S}_p [q_p] \,, \cdots) \,,   \label{2.20}
\end{equation}
where $\bar{S}_v [q_v] \,, \,\, (v = \infty, 2, \cdots , p, \cdots
)$ are given by (\ref{2.17}),  becomes adelic if $T$ is principal
idele and $q'' , \, q'$ are adeles.\vsj

\noi For some other $p$-adic and adelic quadratic classical
cosmological models see \cite{dragovich1}.\vss

\section{{\it\lowercase{$\textbf{p}$}}-Adic and Adelic Quantum Cosmology}

$p$-Adic and adelic quantum cosmology
\cite{arefeva1,dragovich2,dragovich3,dragovich1} is an appropriate
application of $p$-adic \cite{vladimirov2} and adelic quantum
mechanics \cite{dragovich4,dragovich5,dragovich5a} to the universe
as a whole at its very early stage of evolution. Let us now first
recall the basic properties of $p$-adic and adelic quantum
mechanics.

\subsection{$p$-Adic and Adelic Quantum Mechanics}

\noi It is remarkable that ordinary quantum mechanics on a real
space can be generalized to quantum mechanics on $p$-adic spaces for
any prime number $p$. There are two main approaches:  with
complex-valued \cite{vladimirov2} and $p$-adic valued
\cite{khrennikov1} elements of the Hilbert space on ${\mathbb
Q}_p^n$. $p$-Adic quantum mechanics with complex-valued wave
functions is more suitable for connection with ordinary quantum
mechanics, and in the sequel we will briefly review only  this kind
of $p$-adic quantum mechanics.\vsj

\noi When wave functions are complex-valued and arguments are
$p$-adic valued, one cannot construct a direct analog of the
Schr\"odinger equation with a $p$-adic version of Heisenberg
algebra. According to the Weyl  quantization, canonical
noncommutativity in $p$-adic case can be introduced by operators
($h=1$)
\begin{equation}
\hat Q_p(\alpha) \psi_p(x)=\chi_p(-\alpha x)\psi_p(x) , \, \quad
\hat K_p(\beta)\psi_p(x)=\psi_p(x+\beta)  \label{3.1.1}
\end{equation}
which satisfy
\begin{equation}
\hat Q_p(\alpha)\hat K_p(\beta)=\chi_p(\alpha\beta) \hat
K_p(\beta)\hat Q_p(\alpha) , \label{3.1.2}
\end{equation}
where $\chi_p (u) = \exp (2\pi i \{ u \}_p)$ is additive character
on the field ${\mathbb Q}_p$ and $\{ u \}_p$ is the fractional part
of $u \in {\mathbb Q}_p$.\vsj

\noi Let ${\hat x}$ and ${\hat k}$ be operators of position $x$ and
momentum $k$, respectively. Let us define operators $\chi_v (\alpha
{\hat x})$ and  $\chi_v (\beta {\hat k})$ by formulas
\begin{equation}
\chi_v (\alpha {\hat x}) \, \chi_v (a  x) = \chi_v (\alpha { x}) \,
\chi_v (a { x}), \, \quad  \chi_v (\beta {\hat k})\, \chi_v (b { k})
= \chi_v (\beta { k})\, \chi_v (b k) ,  \label{3.1.3}
\end{equation}
where index $v$ denotes real  and any $p$-adic case, and
$\chi_\infty (u) = \exp (- 2 \pi i \, u) $. These operators also act
on a function $\psi_v (x)$ , which has the Fourier transform
${\tilde \psi}(k)$, in the following way:
\begin{equation}
\chi_v (-\alpha {\hat x})\, \psi_v (x) = \chi_v (-\alpha {\hat x})\,
\int \chi_v (- k x)\, {\tilde \psi}(k)\, d^nk = \chi_v(-\alpha x)\,
\psi_v(x) ,  \label{3.1.4}
\end{equation}
\begin{equation}
\chi_v (-\beta {\hat k})\, \psi_v (x) = \int \chi_v (-\beta k)\,
\chi_v (- k x)\, {\tilde \psi}(k)\, d^nk =  \psi_v(x + \beta)  ,
\label{3.1.5}
\end{equation}
where integration in $p$-adic case is with respect to the Haar
measure $dk$ with the properties: $d(k +a) = dk,\, \, d(ak)=|a|_p\,
dk$ and $ \int_{|k|_p\leq 1} dk =1$. Comparing (\ref{3.1.1}) with
(\ref{3.1.4}) and (\ref{3.1.5}) we conclude that $ {\hat
Q}_p(\alpha) = \chi_p(-\alpha {\hat x}) , \,\, {\hat K}_p(\beta) =
\chi_p(- \beta {\hat k})$. Instead of the Heisenberg relations one
has
\begin{equation}
\chi_v (- \alpha_i {\hat x}_i) \, \chi_v (- \beta_j \hat{k}_j) =
\chi_v ( \alpha_i \beta_j \delta_{ij})\, \chi_v (- \beta_j {\hat
k}_j) \, \chi_v (- \alpha_i \hat{x}_i), \label{3.1.6}
\end{equation}
\begin{equation}
\chi_v (- \alpha_i {\hat x}_i) \, \chi_v (- \alpha_j \hat{x}_j) =
\chi_v (- \alpha_j {\hat x}_j) \, \chi_v (- \alpha_i \hat{x}_i),
\label{3.1.7}
\end{equation}
\begin{equation}
\chi_v (- \beta_i {\hat k}_i) \, \chi_v (- \beta_j \hat{k}_j) =
\chi_v (- \beta_j {\hat k}_j) \, \chi_v (- \beta_i \hat{k}_i).
\label{3.1.8}
\end{equation}

 \noi One can introduce the Weyl  operator
\begin{equation}
W_v(\alpha {\hat x}, \beta {\hat k})=\chi_v(\frac{ 1}{ 2} \alpha
\beta)\, \chi_v (- \beta {\hat k}) \, \chi_v(- \alpha {\hat x}),
%\quad z = (\alpha,\beta )\in {\mathbb Q}_v\times {\mathbb Q}_v,
\label{3.1.9}
\end{equation}
which satisfies  relation
\begin{equation}
W_v(\alpha {\hat x}, \beta {\hat k})\, W_v(\alpha' {\hat x}, \beta'
{\hat k}) =\chi_v(\frac{ 1}{ 2}( \alpha \beta' - \alpha' \beta)) \,
W_v((\alpha +\alpha'){\hat x}, (\beta +\beta') {\hat k})
%\quad z = (\alpha,\beta )\in {\mathbb Q}_v\times {\mathbb Q}_v,
\label{3.1.10}
\end{equation}
and is a unitary representation of the Heisenberg-Weyl group. Using
 $W_v(\alpha {\hat x}, \beta {\hat k})$ one obtains generalized
Weyl formula for quantization
\begin{equation}
{\hat f}_v ({\hat k}, {\hat x})= \int W_v (\alpha {\hat x}, \beta
{\hat k}) \, {\tilde f}_v(\alpha, \beta) \, d^n\alpha \, d^n\beta .
\label{3.1.11}
\end{equation}

\noi As a basic instrument to treat dynamics of a $p$-adic quantum
model is natural to take the kernel ${\cal K}_p(x'',t'';x',t')$ of
the evolution operator $U_p(t'',t')$. This kernel obtains by
generalization of its real analog, i.e.
\begin{equation}
\psi_v (x'',t'') = \int {\cal K }_v(x'',t'';x',t')\, \psi_v (x',t')
\, d^nx' \,, \label{3.1.12}
\end{equation}
where ${\cal K}_v(x'',t'';x',t')$ for quadratic Lagrangians can be
defined by path integral
\begin{equation}\label{3.1.13} {\cal K}_v (x'',t'';x',t') =\int_{(x',t')}^{(x'',t'')}
\chi_v ( - \int_{t'}^{t''} L({\dot q}, q, t)\, dt)\, {\cal D}_v q
\,.
\end{equation}\vsj

\noi In the Vladimirov-Volovich formulation \cite{vladimirov2},
$p$-adic quantum mechanics is  a triple
\begin{equation}
(L_2({\mathbb Q}_p), W_p(z), U_p(t)), \label{3.1.14}
\end{equation}
where $W_p (z)$ corresponds to  $W_p (\alpha {\hat x}, \beta {\hat
k})$ defined in \eqref{3.1.9}. \vsj

\noi Adelic quantum mechanics \cite{dragovich2} is a natural
generalization of the above formulation of ordinary and $p$-adic
quantum mechanics: $ (L_2({\mathbb A}), W_{\mathbb A}(z), U_{\mathbb
A}(t))$.\vsj

\noi In complex-valued adelic analysis it is worth mentioning an
additive character
\begin{equation}
 \chi_{\mathbb A} (x) = \chi_\infty (x_\infty) \prod_p \chi_p (x_p),
 \label{3.1.15}
\end{equation}
a multiplicative character
\begin{equation}
  |x|_{\mathbb A}^s = |x_\infty|_\infty^s \prod_p |x_p|_p^s, \ \ s\in {\mathbb C},
                                                    \label{3.1.16}
\end{equation}
and elementary functions of the form
\begin{equation}
 \varphi_{\mathcal P} (x) = \varphi_\infty (x_\infty) \prod_{p\in {\mathcal P}} \varphi_p (x_p)
 \prod_{p\not\in {\mathcal P}} \Omega (|x_p|_p),           \label{3.1.17}
\end{equation}
where $\varphi_\infty (x_\infty)$ is an infinitely differentiable
function on ${\mathbb R}$ and $|x_\infty |_\infty^n \varphi_\infty
(x_\infty) \to 0$ as $|x_\infty|_\infty \to \infty$ for any $n\in
\{0,1,2,\cdots  \}$,  $\varphi_p (x_p)$ are some locally constant
functions with compact support, and
\begin{equation}
\Omega (|x_p|_p) = \left\{  \begin{array}{ll}
                 1,   &   |x_p|_p \leq 1,  \\
                 0,   &   |x_p|_p > 1 .
                 \end{array}    \right.
                 \label{3.1.18}
\end{equation}
All finite linear combinations of elementary functions
(\ref{3.1.17}) make the set $\mathcal{S}({\mathbb A})$ of the
Schwartz-Bruhat adelic functions. The Fourier transform of $\varphi
(x)\in \mathcal{S}({\mathbb A})$, which maps $\mathcal{S}(\mathbb
A)$ onto $\mathcal{S}({\mathbb A})$, is
\begin{equation}
 \tilde{\varphi}(y) = \int_{\mathbb A} \varphi (x)\chi_{\mathbb A} (xy)dx,
 \label{3.1.19}
\end{equation}
where $\chi_{\mathbb A} (xy)$ is defined by (\ref{3.1.15}) and $dx =
dx_\infty \, dx_2 \, dx_3\, \cdots$ is the Haar measure on ${\mathbb
A}$. \vsj

\noi  A basis of $L_2({\mathbb A} (\mathcal P))$ may be given by the
 corresponding
orthonormal eigenfunctions in a spectral problem of the evolution
operator $U_{\mathbb A} (t)$, where $t\in {\mathbb A}$. Such
eigenfunctions have the form
\begin{equation}
\psi_{{\mathcal P}} (x,t) = \psi_{\infty}(x_\infty,t_\infty)
 \prod_{p\in {\mathcal P}} \psi_{p} (x_p,t_p)
 \prod_{p\not\in {\mathcal P}} \Omega (|x_p|_p),
 \label{3.1.20}
\end{equation}
where $\psi_{\infty} \in L_2({\mathbb R})$ and $\psi_{p} \in
L_2({\mathbb Q}_p) $ are eigenfunctions in ordinary and $p$-adic
cases, respectively.  $\Omega (|x_p|_p)$ is an element of
$L_2({\mathbb Q}_p)$, defined by (\ref{3.1.18}), which is invariant
under transformation of an evolution operator $U_p(t_p)$ and
provides convergence of the infinite product (\ref{3.1.20}).
Elements of $L_2({\mathbb A})$ may be regarded as
superpositions\linebreak $\psi (x) = \sum_{{\mathcal P}} C({\mathcal
P})\, \psi_{{\mathcal P}}(x), $ where $\psi_{{\mathcal P}}(x)\in
L_2({\mathbb A}({\mathcal P}))$ (\ref{3.1.20}) and $\sum_{{\mathcal
P}} |C({\mathcal P})|_\infty^2 =1$.\vsj

\noi Theory of $p$-adic generalized functions is presented in
\cite{vladimirov1} and a theory of generalized functions on adelic
spaces is in progress \cite{dragovich6}.\vsj

\noi Adelic evolution operator $U_{\mathbb A}(t)$ is defined by \bea
\nonumber U_{\mathbb A}(t'')\, \psi(x'')&\hspace{-2mm}=\hspace{-2mm}
&\int_{{\mathbb A}} {\mathcal K}_{\mathbb A}(x'',t'';x',t')\,
\psi(x',t')\, dx'\\ &\hspace{-2mm}=\hspace{-2mm}&\prod\limits_{v}{}
\int_{{\mathbb Q}_{v}}{\mathcal K}_v (x''_{v},t''_v;x'_v,t'_v)\,
\psi_v(x'_v,t'_v)\, dx'_v. \label{3.1.21} \eea
 The eigenvalue problem for $U_{\mathbb A}(t)$ reads
\begin{equation} U_{\mathbb A}(t)\, \psi _{{\mathcal P} }
(x)=\chi_{\mathbb A} (E_{\alpha}\, t)\,  \psi _{{\mathcal P}} (x) ,
\label{3.1.22}
\end{equation}
where $\psi_{{\mathcal P} } (x)$ are adelic eigenfunctions
(\ref{3.1.20}), and $E_{\alpha }=(E_{\infty}, E_{2},..., E_{p},...)$
is the corresponding adelic energy.\vsj

\noi Adelic quantum mechanics takes into account ordinary as well as
$p$-adic quantum effects and may be regarded as a starting point for
construction of a more complete quantum cosmology \cite{dragovich2},
quantum field theory \cite{dragovich7} and string/M-theory
\cite{dragovich8,dragovich8a}. In the limit of large distances
adelic quantum mechanics effectively becomes the ordinary one
\cite{dragovich9}.\vsj

\noi Evaluation of  $v$-adic kernel ${\mathcal K}_v(x'',t'';x',t')$
of the unitary evolution operator for one-dimensional systems with
quadratic Lagrangians has the form
\cite{dragovich10,dragovich11,dragovich12}   \bea
\hspace{-10mm}{\mathcal K}_v(x'',t'';x',t')
&\hspace{-2mm}=\hspace{-2mm}& \lambda_v\left( -\frac{1}{2h}
\frac{\partial^2}{\partial x'' \partial
x'}\bar{S}(x'',t'';x',t') \right) \nn \\
&& \times\,\left\vert \frac{1}{h} \frac{\partial^2} {\partial x''
\partial x'}\bar{S}(x'',t'';x',t') \right\vert_v^{\frac{1}{2}}
\chi_v\left( -\frac{1}{h} \bar{S}(x'',t'';x',t')  \right),
                                                      \label{3.1.22}
\eea where $\lambda_v$-functions are presented in
\cite{vladimirov1}.
\bigskip

\subsection{{${p}$}-Adic and adelic wave functions  of the universe}

\noi The universe should be a quantum system, especially at the very
beginning of its evolution. The main task  of quantum cosmology is
to provide formalism to specify quantum states and describe quantum
dynamics of the universe as a whole. The quantum state of the
universe at the beginning is of central importance, since it
determines initial condition for its later behavior. Considering the
universe as a quantum-mechanical system  it has a quantum state
which is encoded in the corresponding wave function.\vsj

\noi This wave function is complex-valued and depends on some real
quantities in standard approach. To include $p$-adic effects one has
to reconsider its formulation. We maintain here the standard point
of view that the wave function takes complex values, but we treat
its arguments (space-time coordinates, gravitational and matter
fields) to be not only real but also $p$-adic and adelic.\vsj

\noi There is not $p$-adic generalization of the Wheeler - De Witt
equation for cosmological models. Instead of differential approach,
Feynman's path integral method in the Hartle-Hawking approach
\cite{hartle} was exploited \cite{dragovich2} and minisuperspace
cosmological models are also investigated by means of adelic quantum
mechanics \cite{dragovich1}.\vsj

\noi $p$-Adic and  adelic minisuperspace quantum cosmology is an
application of $p$-adic and adelic quantum mechanics to the
cosmological models, respectively. In the path integral approach to
standard quantum cosmology, the starting point is Feynman's path
integral method. The amplitude to go from one state with intrinsic
metric $h_{ij}'$ and matter configuration $\phi'$ on an initial
hypersurface $\Sigma'$ to another state with metric $h_{ij}''$ and
matter configuration $\phi''$ on a final hypersurface $\Sigma''$ is
given by the path integral
\begin{equation}
{\mathcal K}_\infty ( h_{ij}'',\phi'',\Sigma'';
h_{ij}',\phi',\Sigma') = \int
\chi_\infty(-S_\infty[g_{\mu\nu},\Phi])\, \, {\cal
D}_\infty{g_{\mu\nu}} \, {\cal D}_\infty\Phi  \label{3.2.1}
\end{equation}
over all four-geometries $g_{\mu\nu}$ and matter configurations
$\Phi$, which interpolate between the initial and final
configurations. In (\ref{3.2.1}) $S_\infty [g_{\mu\nu},\Phi]$ is an
Einstein-Hilbert action for the gravitational and matter fields. To
perform $p$-adic and adelic generalization we make first $p$-adic
counterpart of the action using form-invariance under change of real
to the $p$-adic number fields. Then we generalize (\ref{3.2.1}) and
introduce $p$-adic complex-valued cosmological amplitude
\begin{equation}
 {\mathcal K}_p ( h_{ij}'',\phi'',\Sigma''; h_{ij}',\phi',\Sigma') =
 \int
\chi_p(-S_p[g_{\mu\nu},\Phi]) \, \, {\cal D}_p{g_{\mu\nu}} \, {\cal
D}_p\Phi. \label{3.2.2}
\end{equation}

\noi The standard minisuperspace ground-state wave function in the
Hartle-Hawking (no-boundary) proposal \cite{hartle} is defined by
functional integration in the Euclidean version of
\begin{equation}
\psi_\infty[h_{ij}]= \int \chi_\infty(-S_\infty[g_{\mu\nu},\Phi]) \,
\, {\cal D}_\infty g_{\mu\nu} \, {\cal D}_\infty\Phi, \label{3.2.3}
\end{equation}
over all compact four-geometries $g_{\mu\nu}$ which induce $h_{ij}$
at the compact three-manifold. This three-manifold is the only
boundary of the all four-manifolds. Extending Hartle-Hawking
proposal to the $p$-adic minisuperspace  \cite{arefeva1},  an adelic
Hartle-Hawking wave function is the infinite product
\begin{equation}
\psi_{\mathbb A}(q)= \prod_{v}\int  \chi_v(-S_v [g_{\mu\nu},\Phi])
\, \, {\cal D}_vg_{\mu\nu} \, {\cal D}_v\Phi , \label{3.2.4}
\end{equation}
where path integration must be performed over both, Archimedean and
non-Archimedean geometries. If an  evaluation of the corresponding
functional integrals for a minisuperspace model yields
$\psi(q_\alpha)$ in the form (\ref{3.1.20}), then  such cosmological
model is a Hartle-Hawking adelic one.\vsj

\noi Before to proceed in the above way it is worth mentioning
another approach \cite{arefeva1} which was the first one in $p$-adic
quantum cosmology. The essence of this approach consists in the
following $p$-adic proposal for the Hartle-Hawking type of the wave
function:
\begin{equation}
\psi_\infty (q) = \sum_{a.m.} \prod_{p}\int  \chi_p(-S_p
[g_{\mu\nu},\Phi]) \, \, {\cal D}_p g_{\mu\nu} \, {\cal D}_p\Phi \,,
\label{3.2.5}
\end{equation}
where summation is over algebraic manifolds. This proposal was
illustrated on the above de Sitter model with the Euclidean version
of $ds^2$ in \eqref{2.13}.\vsj

\noi It is shown \cite{dragovich2} that the de Sitter minisuperspace
model in $D=4$ space-time dimensions is  the Hartle-Hawking adelic
one. Namely, according to the Hartle-Hawking proposal one has
\begin{equation}
\psi_v (q) = \int {\mathcal K}_v (q, T; 0, 0)\, dT \,, \quad v =
\infty, 2, 3, \cdots, p , \cdots \,,   \label{3.2.6}
\end{equation}
where
\begin{equation} {\mathcal K}_v (q'', T; q', 0) = \lambda_v (- 8 T)
|4 T|_v^{-\frac{1}{2}}  \chi_v \Big[- \frac{\lambda^2 T^3}{24} +
(\lambda q - 2) \frac{T}{4} + \frac{q^2}{8 T}\Big] \label{3.2.7}
\end{equation}
is the kernel of the $v$-adic evolution operator. The functions
$\lambda_v (a)$ have the  properties \cite{vladimirov1}
\begin{equation}
|\lambda_v (a)|_v = 1\,, \,\, \lambda_v (b^2 a) =\lambda_v (a)\,,
\lambda_v (a) \, \lambda_v(b) = \lambda_v (a + b) \, \lambda_v (a b
(a+b)) \,.  \label{3.2.8}
\end{equation}
Employing the $p$-adic Gauss integral \cite{vladimirov1}
\begin{equation}
\int_{\mathbb{Q}_p} \chi_p (\alpha x^2 + \beta x) \, dx = \lambda_p
(\alpha) |2 \alpha|_p^{-\frac{1}{2}} \, \chi_p \Big( -
\frac{\beta^2}{4 \alpha} \Big) \,, \quad \alpha\neq 0 \,,
\label{3.2.9}
\end{equation}
one can rewrite $p$-adic version of \eqref{3.2.6} in the form
\begin{equation}
\psi_p (q)  = \int_{\mathbb{Q}_p} dx \, \chi_p (q x) \, \int DT
\chi_p \Big[ -\frac{\lambda^2 T^3}{24} + \Big( \frac{\lambda q}{4} -
\frac{1}{2} - 2 x^2\Big) T \Big] \,. \label{3.2.10}
\end{equation}
Taking the region of integration to be $|T|_p \leq 1$ one obtains
\begin{equation}
\psi_p (q) = \int_{\mathbb{Q}_p} dx \, \chi_p (q x) \, \Omega \big (
 \big| \frac{\lambda q}{4} - \frac{1}{2} - 2 x^2 \big|_p  \big) \,,
 \quad \big| \frac{\lambda^2}{24} \big|_p \leq 1 \,. \label{3.2.11}
\end{equation}

\noi An evaluation of the integral \eqref{3.2.11} yields
\begin{equation}
\psi_p (q) = \exp (i \pi\, \delta^1_{|q|_2} \, \delta_p^2) \, \Omega
(|q|_p) \,, \quad \big| \frac{\lambda^2}{24} \big|_p \leq 1 \,,
\label{3.2.12}
\end{equation}
where $\delta_a^b$ is the Kronecker symbol. $\psi_\infty (q_\infty)$
was explored in \cite{halliwell} and the result depends on the
contour of integration and has an exact solution
\begin{equation}
\psi_\infty (q_\infty) = \exp\big( \frac{1}{3 \lambda} \big) \, Ai
\Big( \frac{1 - \lambda\, q_\infty}{ (2 \lambda)^{\frac{2}{3}}}
\Big)  \,,   \label{3.2.13}
\end{equation}
where $Ai (x)$ is the Airy function. Finally we obtain an adelic
wave function for the de Sitter cosmological model in the form
\begin{equation}
\psi_{\mathbb A} (q) = \psi_\infty (q_\infty) \, \prod_p \exp (i \pi
\, \delta^1_{|q|_2} \, \delta_p^2) \,\, \Omega (|q_p|_p) \,, \quad
\big| \frac{\lambda^2}{24} \big|_p \leq 1 \,. \label{3.2.14}
\end{equation}

\noi It is shown in \cite{dragovich3,dragovich1}   that $p$-adic and
adelic generalization of the minisuperspace cosmological models can
be successfully performed in the framework of $p$-adic and adelic
quantum mechanics \cite{dragovich4,dragovich5} without use of the
Hartle-Hawking approach.  The following cosmological models are
investigated \cite{dragovich1}: the de Sitter model, model with a
homogeneous scalar field, anisotropic Bianchi model with three scale
factors and some two-dimensional minisuperspace models.The necessary
condition that a system can be regarded as the adelic one is the
existence of $p$-adic ground state $\Omega (|q_\alpha|_p)  \,\,\,$ $
(\alpha = 1, 2, \cdots , n)$ in the way
\begin{equation}
\int_{|q'_\alpha|_p \leq 1} \mathcal{K}_p (q^{''}_\alpha \,, T ;
q^{'}_\alpha \,, 0 ) \, dq'_\alpha  = \Omega (|q^{''}_\alpha|_p)
\label{3.2.15}
\end{equation}
 for all $p$ but a finite set $\mathcal{P}$.
For the case of de Sitter model one obtains
\begin{equation}
\psi_p (q) = \left\{  \begin{array}{ll}
                 \Omega (|q|_p)\,, \quad |T|_p \leq 1 \,, & \big| \frac{\lambda^2}{24} \big|_p \leq 1 \,, \quad p \neq 2 \,,  \\
                 \Omega (|q|_2)\,, \quad |T|_2 \leq \frac{1}{2} \,, & \big| \frac{\lambda^2}{24} \big|_2 \leq 1 \,, \quad p=2
                 \,,
                 \end{array}    \right.
                 \label{3.2.16}
\end{equation}
what is in a good agreement with the result \eqref{3.2.14} obtained
by the Hartle-Hawking proposal. Some other forms of the $p$-adic
wave functions are also found \cite{dragovich1}, e.g. $\psi_p (q) =
\Omega ( p^\mu |q|_p)$ and $\psi_p (q) = \delta (|q|_p  - p^\mu)$
where $\mu \in \mathbb{Z}$.\vsj

\noi Let us now consider the interpretation of an adelic wave
function. Note that there is a general problem of the usual
interpretation of the wave function of the universe. Here we are
going to discuss only adelic aspects. It is natural to investigate
adelic wave function on the rational values of its arguments.
Without loss of generality we can use the above de Sitter
cosmological model. For the adelic ground state \eqref{3.2.14} the
density of probability in rational points is
\begin{equation}
|\psi_{\mathbb A} (q)|_\infty^2 = |\psi_\infty (q) |^2_\infty \,
\prod_p \Omega (|q|_p) = \left\{  \begin{array}{ll}
                 |\psi_\infty (q) |^2_\infty \,, &  q \in \mathbb{Z} \,,  \\
                 0 \,, &  q \in \mathbb{Q} \setminus \mathbb{Z}
                 \,.
                 \end{array}    \right.
                 \label{3.2.17}
\end{equation}

\noi When (the density of) probability is equal (or close) to unity
or zero it can be regarded as a certain event. From \eqref{3.2.17}
it follows that $q$-space is discrete, because the density
probability is nonzero only at integers. Note that here spacing is
related to the Planck length but in some other adelic quantum models
it will correspond to the characteristic length of the system. If we
integrate \eqref{3.2.14} over $|q_p|_p \leq 1$ for every $p$ then we
obtain ordinary wave function. In the case of adelic wave function
of the form \eqref{3.1.20} with $\mathcal P \neq \emptyset$,
$q$-space has not a sharp discreteness but fuzzyness.\vsj

\noi As a result of $p$-adic effects,  adelic quantum systems
contain discreteness of the space with spacing equal to the own
characteristic length. This kind of discreteness was obtained for
the first time in the context of the Hartle-Hawking adelic de Sitter
quantum model \cite{dragovich2}. At the distances much larger than
the characteristic length instead of an adelic wave function it is
enough to use only the ordinary one over real space. \vss

\section{Real and\, {\it\lowercase{$\textbf{p}$}}-Adic Worlds. \, Adelic Universe}

\subsection{On Adelic Models}

\noi Let us introduce two distinct types of adelic models: ({\it i})
principal adelic models and ({\it ii}) non-principal adelic
models.\vsj

\noi In principal adelic models all parameters and constants, which
characterize physical system, are principal ideles, i.e.  they are
nonzero rational numbers which are the same in real and all $p$-adic
counterparts.  Adelic product formulas can exist in these models,
which connect real and $p$-adic aspects of the same quantity. A
simple and illustrative such formula is
\begin{equation}
|x|_\infty^c \, \prod_p |x|_p^c  =1\,, \quad  c\in \mathbb{C}\,,
\quad x \in \mathbb{Q}\setminus \{ 0\} \,, \label{4.1}
\end{equation}
which is adelic multiplicative character on principal ideles.
Another simple example is adelic additive character  on principal
adeles, i.e.
\begin{equation}
\chi_\infty (x)\, \prod_p \chi_p (x) = 1 \,, \quad x \in
\mathbb{Q}\,. \label{4.2}
\end{equation}
Employing $|x|_v^{\alpha -1}$ and $\chi_v (x)$ in some integrals one
can obtain new product formulas. For instance (see
\cite{vladimirov3} and references therein)
\begin{equation}
\Gamma_\infty (\alpha) \, \prod_p \Gamma_p(\alpha) = 1 \,, \quad
\Gamma_v (\alpha) = \int_{\mathbb{Q}_v} |x|_v^{\alpha - 1}\, \chi_v
(x) \, d_vx \,, \quad \alpha \neq 0, 1 \,, \label{4.3}
\end{equation}
\begin{equation}
B_\infty (a, b)\, \prod_p B_p (a, b) = 1 \,,  \quad B_v (a, b) =
\int_{\mathbb{Q}_v} |x|_v^{a-1} \, |1 - x|_v^{b-1} d_vx \,, \quad a
+ b + c = 1 \,.  \label{4.4}
\end{equation}
While the formulas \eqref{4.1} and \eqref{4.2} are valid only for
rational numbers, \eqref{4.3} and \eqref{4.4} have place for real
and complex numbers. The product formula \eqref{4.4} connects real
and $p$-adic amplitudes for scattering of two open string tachyons
\cite{freund1}. As a result of \eqref{4.4}, complicated real
amplitude  can be expressed as product of inverse $p$-adic
counterparts which are elementary functions. There are also some
product formulas which include all $p$-adic counterparts but not the
real one. Such an example is
\begin{equation}
 \prod_p \Omega (|x|_p) = \left\{
\begin{array}{ll}
                 1 \,, &  x \in \mathbb{Z} \,,  \\
                 0 \,, &  x \in \mathbb{Q} \setminus \mathbb{Z}
                 \,.
                 \end{array}    \right.
                 \label{4.5}
\end{equation}
We used \eqref{4.5} in \eqref{3.2.17}. \vsj

\noi Adelic quantum mechanics \cite{dragovich4,dragovich5} and its
generalizations to quantum field theory \cite{dragovich7} and string
theory \cite{dragovich8} belong to the principal adelic models.\vsj

\noi In non-principal adelic models parameters and constants are
adeles but not principal ones, i.e. they are not rational numbers
the same for real and $p$-adic counterparts.  A non-principal adelic
constant (or parameter) is an adele
\begin{equation}
c_{\mathbb{A}} = (c_\infty ,\, c_2 , \, c_3 , \,\cdots , c_p ,\,
\cdots) \,, \label{4.6}
\end{equation}
where at least two of components $c_v$ are not the same rational
number. There are infinitely many possibilities to fix all  $c_v$ in
\eqref{4.6}. One of them is the case when $c_p = p$ for all primes
$p$. Another interesting case is when all but a finite number of
$c_v$ are equal zero. In particular, all but one of $c_v$ can be
zero. If only $c_\infty \neq 0$, adelic model is reduced to the real
one. Analogously if  only $c_p \neq 0 $ for a fixed $p$, we have a
p-adic model. In the sequel we will pay some attention to the case
when $c_\infty \neq 0$ and $c_p  \neq 0$ for a finite set $\mathcal
P$ of primes $p$.

\subsection{$p$-Adic Matter, Dark Energy and Dark Matter}

\noi Applying classical Einstein equations \eqref{2.7} and the FRW
metric \eqref{2.12} (with $N=1, \,\, c=1$) to the universe, one
obtains
\begin{equation}
\frac{\ddot{a}}{a} = - \frac{4 \pi G}{3} \sum_i (\rho_i  + 3 p_i) +
\frac{\Lambda}{3} \,, \quad \Big( \frac{\dot{a}}{a} \Big)^2  +
\frac{k}{a^2} = \frac{8 \pi G}{3 }\, \sum_i \rho_i  +
\frac{\Lambda}{3}\,,  \label{4.2.1}
\end{equation}
where $\rho_i$ and $p_i$ are $i$-th component of energy and pressure
density, respectively, and $a$ is cosmological scale factor. The
equation of state is $p_i = w_i \,\rho_i$. Since $\Lambda$ can be
regarded as an energy density we can rewrite equations \eqref{4.2.1}
in the following form
\begin{equation}
\frac{\ddot{a}}{a} = - \frac{4 \pi G}{3} \sum_i (\rho_i  + 3 p_i)
\,,   \label{4.2.2}
\end{equation}
\begin{equation}
 \Big( \frac{\dot{a}}{a} \Big)^2  +
\frac{k}{a^2} = \frac{8 \pi G}{3 }\, \sum_i \rho_i  \,.
\label{4.2.3}
\end{equation}
Recent observations (see reviews \cite{sahni1,padmanabhan1}) show
that
\begin{equation}
\Omega = \sum_i \frac{\rho_i}{\rho_{cr}} = \sum_i \Omega_i \approx
1\,, \quad \Omega_B \approx 0.04 \,, \quad \Omega_{DM} \approx 0.26
\,, \quad \Omega_{DE} \approx 0.70 \,, \label{4.2.4}
\end{equation}
where $\Omega_B \,, \,\, \Omega_{DM} \, $ and $\Omega_{DE}$ are
related to baryonic matter, dark matter and dark energy,
respectively, and $\rho_{cr} = \frac{3}{8\, \pi\, G}\, H_0^2 $ is
critical energy density required for the universe to be spatially
flat. $H_0 = \big( \frac{\dot{a}}{a} \big)_0$ is the rate of
expansion at present time. While baryonic matter is partially
luminous, the nature of dark matter and dark energy is presently
unknown.\vsj

\noi Dark matter was introduced (already in 1930's) to explain
dynamics of galaxies in clusters as well as dynamics inside
individual galaxies. It is nonbaryonic and clustered (attractive)
matter. Many nonbaryonic particles (in particular, neutralino and
axion) have been considered as candidates for dark matter.\vsj

\noi Dark energy was introduced as an unclustered (repulsive) matter
with negative pressure and responsible for accelerated expansion of
the universe. Cosmic acceleration was discovered recently
\cite{perlmutter,riess} in observations of cosmologically distant
Type Ia Supernovae (SNeIa) and independently verified by Cosmic
Microwave Background and large scale structure investigations.
Namely, to have an acceleration ($\ddot{a} >0$), the RHS of eq.
\eqref{4.2.2} has to be positive, i.e. $\sum_i (\rho_i + 3 p_i) <
0$. Since $\rho_B + 3 p_B > 0$ and $\rho_{DM} + 3 p_{DM} > 0$ it
follows that $\rho_{DE} + 3 p_{DE} < 0$. Positivity of $\rho_{DE}$
yields $p_{DE} < - \frac{\rho_{DE}}{3} < 0$, i.e. $w < -1/3$.
Current observational value of parameter $w$ (see \cite{sahni1} and
references therein) is in the range $-1.61 < w < -0.78$.\vsj

\noi There are many candidate models for the dark energy. The
simplest model is the cosmological constant with $w = -1 \quad (p =
- \rho = const. )$. Some other dynamical scalar field  models are
\cite{sahni1}: quintessence, braneworld, Chaplygin gas and Phantom
dark energy (with $w < -1$). In spite of many interesting results,
neither these nor other so far proposed models offer satisfactory
answers to all questions around the dark energy.\vsj

\noi I propose here $p$-adic origin of the dark energy and dark
matter. In other words, dark energy and dark matter are two forms of
the $p$-adic matter. If we take string theory seriously then there
exist not only real strings but also $p$-adic strings. Real and
$p$-adic matter  consist of real and $p$-adic strings, respectively.
Even without string theory, using $p$-adic quantum-mechanical models
one can argue possible existence of $p$-adic matter.\vsj

\noi An instructive model of $p$-adic matter is nonlocal and
nonlinear effective scalar field theory \cite{brekke1,frampton1}
which describes dynamics of $p$-adic tachyonic  matter. The
corresponding action is given by
\begin{equation}
S = \int d^dx \,\, \mathcal{L}_p  \,, \quad \mathcal{L}_p  =
\frac{1}{g_p^2} \Big[ -\frac{1}{2}\phi\, p^{- \frac{1}{2}\,\Box} \,
\phi + \frac{1}{p+1}\, \phi^{p + 1} \Big] \,, \label{4.2.5}
\end{equation}
where
\begin{equation}
\frac{1}{g_p^2} = \frac{1}{g^2} \, \frac{p^2}{p-1} \,, \quad \Box =
- \frac{\partial^2}{\partial t^2} +  \frac{\partial^2}{\partial
x_1^2} + \cdots + \frac{\partial^2}{\partial x_{d-1}^2} \,,
\label{4.2.6}
\end{equation}\vsj

\noi and the equation of motion is
\begin{equation}
p^{- \frac{1}{2}\,\Box} \, \phi \, =  \, \phi^{p} \,.  \label{4.2.7}
\end{equation}
This $p$-adic string theory was employed in the context of tachyon
condensation \cite{sen1} and for a recent review see \cite{freund2}.
Some mathematical aspects of time dependent solutions of
\eqref{4.2.7} were analyzed in \cite{vladimirov4} (see also
references therein). It is worth mentioning that this $p$-adic
tachyon theory resembles dynamics of nonlocal real tachyon in string
field theory, which is also under consideration as a candidate for
cosmological dark energy \cite{arefeva2}. It would be of great
importance to find $p$-adic valued counterpart of Lagrangian
\eqref{4.2.5} which reproduces the same complex-valued tachyon
amplitudes at the tree level,  as well as to establish a map between
them. \vsj

\noi According to this proposal baryonic matter is real one, while
dark energy and dark matter have $p$-adic origin. Work on a concrete
$p$-adic model of dark energy and dark matter is in progress.

\subsection{Adelic Universe with Real and  $p$-Adic Worlds }

\noi An example of a non-principal adelic physical system could be
the universe as a whole. Namely, it seems natural to regard our
universe as adelic one consisting of real and $p$-adic worlds which
interact only gravitationally. Each of these worlds is made of its
own matter and their main characteristics can be described by
relevant real or $p$-adic numbers.\vsj

\noi Even if we suppose that at the very beginning the universe was
as a principal adelic system it could evolve to the non-principal
one. Namely, according to string landscape (see recent discussion in
\cite{weinberg}), effective potential in string theory has many
local maxima and minima which determine values of various
parameters. During evolution they can lead to different values of
parameters in real and $p$-adic sectors and definitely to
transformation of principal adelic strings to real and $p$-adic
ones, which may be regarded as non-principal adelic strings.

\vss

\section{Concluding Remarks}

\noi In an introductory way we have reviewed basic properties of
$p$-adic and adelic cosmology. Proposal on $p$-adic origin of dark
energy and dark matter, as well as on $p$-adic worlds of the
universe, has been put forward.\vsj

\noi It is worth mentioning some coincidences between $p$-adics and
dimensionality of the space-time. A nonsingular form of an odd
degree  $r$ \cite{shafarevich}
\begin{equation}
\sum_{i=1}^n \, a_i \, x_i^r \,, \quad a_i \in \mathbb{Q}  \,, \quad
r = 2 k + 1 \,, \quad k \in \mathbb{N} \cup \{ 0\} \,, \label{5.1}
\end{equation}
represents all $p$-adic numbers in a nontrivial way in
$\mathbb{Q}_p$ if $n \geq D = r^2 + 1$. Also the above form
\eqref{5.1} nontrivially  represents all rational numbers in
$\mathbb{Q}$ when $n \geq D = r^2 + 1$. It follows that the lowest
dimension $D$ for $r = 3$ is $D= 10$ and for $r = 5 \,, \,\, D =
26$. It  was noted \cite{dragovich13} that these dimensions $D$
coincide with critical dimensions of space-time in string theory,
i.e. $D = 10$ for superstrings and $D = 26$ for bosonic strings.\vsj

\noi Another intriguing coincidence is related to quadratic
algebraic extensions (of $\mathbb{R}$ and $\mathbb{Q}_p$) and
dimensionality of M-theory. Recall \cite{vladimirov1} that in real,
$p$-adic ($p \neq 2$) and $2$-adic cases we have one, three and
seven quadratic extensions, respectively. Their sum gives $1  + 3 +
7 = 11$, what is just number of space-time dimensions in
M-theory.\vsj

\noi $p$-Adic pseudoconstants \cite{mahler} are $p$-adic valued
functions $c_p (t)$ which have derivatives vanishing, i.e.
$\dot{c}_p (t) = 0$ for $t \in \mathbb{Z}_p$. They give new
cosmological possibilities comparing with real analysis and could be
related \cite{arefeva1} to stochastic inflation.

%%%%%%%%%%%%%%%%%%%%%%%%%%%%%%%%%%%%%%%%%%%%
%% Sample figure:
%%
%% The option [height=...] scales the picture to the given height,
%% without it it would be printed at its nominal size
%%%%%%%%%%%%%%%%%%%%%%%%%%%%%%%%%%%%%%%%%%%%

%\begin{figure}
 % \includegraphics[height=.3\textheight]{golfer}
  %\caption{Picture to fixed height}
  %\caption{Picture to fixed height}
%\end{figure}

%%%%%%%%%%%%%%%%%%%%%%%%%%%%%%%%%%%%%%%%%%%%
%% SAMPLE TABLE
%%
%% Shows the use of \tablehead and \tablenote
%% macros
%%%%%%%%%%%%%%%%%%%%%%%%%%%%%%%%%%%%%%%%%%%%

%%%%%%%%%%%%%%%%%%%%%%%%%%%%%%%%%%%%%%%%%%%%%%%%
%% BACKMATTER
%%%%%%%%%%%%%%%%%%%%%%%%%%%%%%%%%%%%%%%%%%%%%%%%
\vsa

\begin{theacknowledgments}
  The work on this article has been supported by the Ministry of
  Science and Environmental Protection of the  Republic of Serbia
  under projects No 1426 and No 144032. I would like to thank I.Ya.
  Aref'eva, V.S. Vladimirov and I.V. Volovich for inspirative and useful
  discussions I have had with them in various periods of work
  on some problems contained in this paper.
\end{theacknowledgments}



\bibliographystyle{aipproc}   % if natbib is available

%\bibliography{sample}

 %{\typeout{}
  %\typeout{******************************************}
  %\typeout{** Please run "bibtex \jobname" to optain}
  %\typeout{** the bibliography and then re-run LaTeX}
  %\typeout{** twice to fix the references!}
 % \typeout{******************************************}
  %\typeout{}
 %}

%\end{document}

%%%%%%%%%%%%%%%%%%%%%%%%%%%%%%%%%%%%%%%%%%%
%% The following lines show an example how to produce a bibliography
%% without the help of the BibTeX program. This could be used instead
%% of the above.
%%%%%%%%%%%%%%%%%%%%%%%%%%%%%%%%%%%%%%%%%%%
\vsr

%\endinput
%%
%% End of file `template-6s.tex'.
\end{document}